\documentclass[sigconf]{acmart}
\usepackage[american]{babel}
\usepackage{multirow}
\usepackage{enumitem}
\usepackage[american]{babel}
\usepackage{microtype}
\usepackage{url}
\AtBeginDocument{%
  \providecommand\BibTeX{{%
    \normalfont B\kern-0.5em{\scshape i\kern-0.25em b}\kern-0.8em\TeX}}}

\copyrightyear{2020}
\acmYear{2020}
\setcopyright{acmcopyright}
\acmConference[MM '20] {28th ACM International Conference on Multimedia}{October 12--16, 2020}{Seattle, WA, USA.}
\acmBooktitle{28th ACM International Conference on Multimedia (MM '20), October 12--16, 2020, Seattle, WA, USA.}
\acmPrice{15.00}
\acmDOI{10.1145/3394171.3414537}
\acmISBN{978-1-4503-7988-5/20/10}

\begin{document}
\fancyhead{}

\title{PyRetri: A PyTorch-based Library for Unsupervised Image Retrieval by Deep Convolutional Neural Networks}


\author{Benyi Hu}
\authornote{This work was made when B. Hu was an intern in Megvii Research Nanjing. This work was supported by the National Key R\&D Program of China (Grant 
2018AAA0102504) and the National Natural Science Foundation of China (Grant 61973245).}
\affiliation{%
  \institution{Xi'an Jiaotong University}
}
\email{hby0906@stu.xjtu.edu.cn}

\author{Ren-Jie Song}
\affiliation{
  \institution{Megvii Technology}
}
\email{songrenjie@megvii.com}

\author{Xiu-Shen Wei}
\authornote{X.-S. Wei (corresponding author) is with PCA Lab, Key Lab of Intelligent Perception and Systems for High-Dimensional Information of Ministry of Education, and Jiangsu Key Lab of Image and Video Understanding for Social Security, School of Computer Science and Engineering, Nanjing University of Science and Technology.}
\affiliation{
  \institution{Nanjing University of Science and Technology}
}
\email{weixs.gm@gmail.com}

\author{Yazhou Yao}
\affiliation{
  \institution{Nanjing University of Science and Technology}
}
\email{yazhou.yao@njust.edu.cn}

\author{Xian-Sheng Hua}
\affiliation{
  \institution{Alibaba Group}
}
\email{huaxiansheng@gmail.com}

\author{Yuehu Liu}
\affiliation{%
  \institution{Xi'an Jiaotong University}
}
\email{liuyh@mail.xjtu.edu.cn}

\renewcommand{\shortauthors}{Hu et al.}

\begin{abstract}
Despite significant progress of applying deep learning methods to the field of content-based image retrieval, there has not been a software library that covers these methods in a unified manner. In order to fill this gap, we introduce PyRetri, an open source library for deep learning based unsupervised image retrieval. The library encapsulates the retrieval process in several stages and provides functionality that covers various prominent methods for each stage. The idea underlying its design is to provide a unified platform for deep learning based image retrieval research, with high usability and extensibility. The project source code, with usage examples, sample data and pre-trained models are available at \url{https://github.com/PyRetri/}.
\end{abstract}

\ccsdesc[500]{Software and its engineering~Software libraries and repositories}
\ccsdesc[300]{Information systems~Top-k retrieval in databases}

\keywords{Open Source, Image Retrieval, Convolutional Neural Networks}


\maketitle

\section{Introduction}

Content-based image retrieval (CBIR), which makes use of the representation of visual content to identify relevant images, is one of the fundamental research challenges extensively studied in the multimedia community for decades~\cite{cbirsurvey}. Recently, with the prosperity of applying deep learning methods, CBIR has also witnessed the prominence of powerful features of convolutional neural networks. However, the pipeline of deep learning based unsupervised image retrieval is complicated, and empirical configurations used in each stage can have a significant impact on retrieval accuracy. Although there are a wide range of open source implementations released by researchers, codes are not organized in some standardized manners, which may be tedious or confusing for users. Therefore, a high quality and unified framework is essential to keep up the rapid pace of innovation for deep learning based image retrieval researches. 

In order to fill this gap, we propose the PyRetri library, an open-source framework that divides deep learning based unsupervised CBIR into several main stages with clear application programming interfaces (APIs). It is a modular Python library based on PyTorch~\cite{paszke2019pytorch} that provides easy-to-use modules to facilitate researchers and engineers in developing unsupervised image retrieval approaches. To the best of our knowledge, this is the first open-source library for unsupervised image retrieval by deep learning.

\begin{figure*}[t]
  \centering
\vspace{-0em}
  \includegraphics[width=0.8\linewidth]{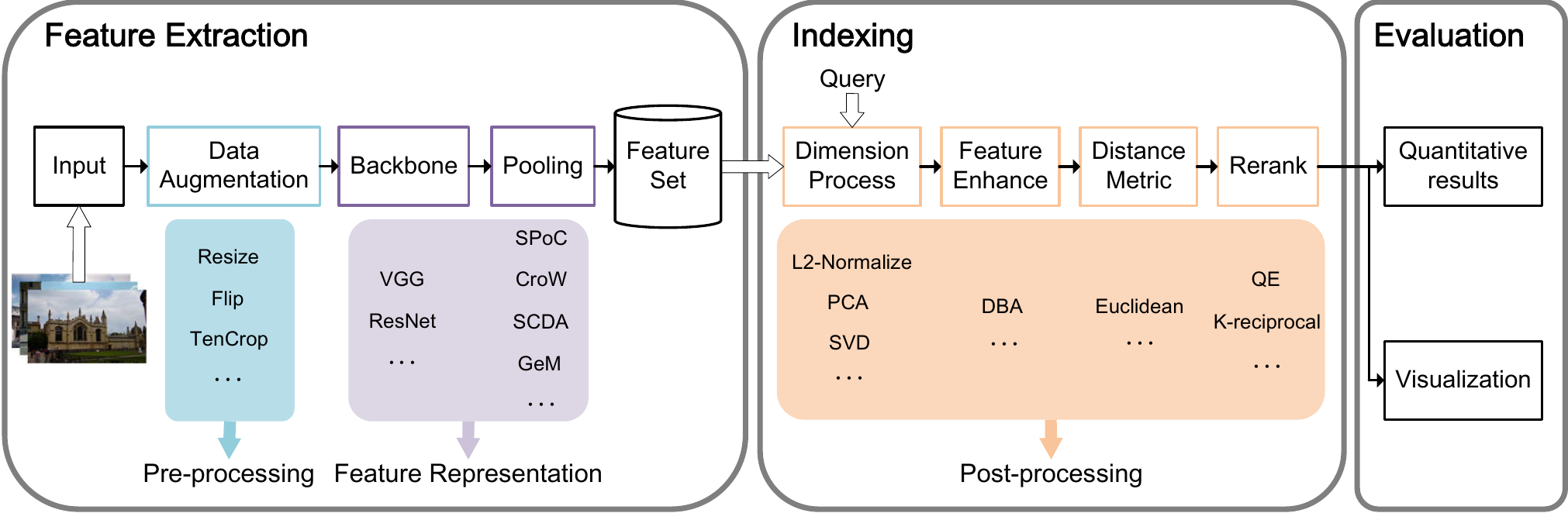}
\vspace{-1em}
  \caption{Framework of deep learning based unsupervised image retrieval, illustrated with abstractions in PyRetri.}\label{fig:overview}
\end{figure*}

Towards the goal of providing a high-quality, easy-to-use and easy-to-extend framework, we obey the following principles: (1) \textbf{High code quality}. We assure the quality by developing through the peer review process and good software engineering practices. For the reason that generality should not come at the cost of usability, we tackle project maintenance difficulties by providing type hints. Furthermore, the library code is maintained in a clean, consistent style, with class and function names that are descriptive of the underlying functionality. (2) \textbf{Human readable configurations}. We utilize \texttt{YACS}~\cite{yacs}, a highly human readable configurations management system, to define all the hyper-parameters of methods used in the CBIR pipeline in only one \texttt{config} file. Through this serialization format, users can easily manage their own experiments with brief and clear configurations. (3) \textbf{Modular design}. Approaches contained in PyRetri do not intend to create module lock-in. Instead, modules are modeled minimal single-function blocks that share an interaction interface, which allows easy plug-ins of user-defined modules. This spawns a unified environment where developers are able to efficiently explore ideas through high-level module operations, and apply customizations to modules only if necessary.

To summarize, the main contributions of PyRetri are:
\begin{itemize}[itemsep=0em, leftmargin=2em]
\item We propose the first open source framework to unify the pipeline of deep learning based unsupervised image retrieval, which is readable and extendable.
\item We provide high quality implementations of CBIR algorithms to solve retrieval tasks with emphasis on usability.
\item We release reference codes, model zoos and tools of instance-level image retrieval, which can benefit researchers to implement and design their own methods.
\end{itemize}

\section{Design Overview}

The overall architecture of PyRetri is illustrated in Figure~\ref{fig:overview}. In PyRetri, the pipeline of deep learning based unsupervised CBIR is grouped into three crucial modules: feature extraction, indexing and evaluation. In the following, we elaborate these modules.

\subsection{Feature Extraction Module}

The feature extraction module is utilized to compute the global-level image representation for retrieval with one single network pass. For practical convenience, we first generate a \texttt{json} file to describe the query or gallery datasets, by saving the information of each image such as its path and labels in a list of dictionaries. Given the data augmentation operations and pre-trained models, PyTorch~\cite{paszke2019pytorch} is adopted as the backend and the inference engine to construct a feature extraction pipeline. The output feature is flexibly assigned through a hooking mechanism, whether it is generated by the fully-connected layers or the convolutional layers. Particularly for the convolutional layers, deep descriptors are firstly collected and then aggregated into the global-level representation, which is necessary and discriminative for CBIR~\cite{wei2017selective}. 

\subsection{Indexing Module}

The indexing module is the core brick of CBIR, which returns images containing the same content as the query based on the similarity between their image representations. The indexing stages of many retrieval tasks share a similar workflow, where query features and gallery features are projected into a new manifold space and distances are calculated between them. As shown in Figure~\ref{fig:overview}, we constructed a complete and meticulous indexing pipeline where interfaces are reserved for all indexing stages. Thanks to the modular design, users can plug in their methods easily through these interfaces without changing the core code of the retrieval pipeline.

Similar to the feature extraction module, the retrieval results of each query, a.k.a. neighboring indexes, are add to the \texttt{json} file, which is convenient for the following evaluation process.

\subsection{Evaluation Module}

The evaluation module is utilized to evaluate the retrieval accuracy and further analyze the retrieval results. We adopt the recall and mean average precision (mAP) as the evaluation metrics, which are widely used in CBIR related tasks. In general, ``recall'' denotes the ratio of returned true matches to the total number or true matches in the database. ``mAP'' denotes the average of AP on all queries, which amounts area under the precision-recall curve. Typically, higher recalls and mAPs mean better retrieval accuracy. In our implementations, we provide interfaces for both content-based image retrieval and person re-identification tasks.

In addition, with clear interfaces, we support visualizing retrieval results of a single query image by showing or saving its top-$k$ returned images, which is convenient for failure case analyses.

\section{Supported Methods}

PyRetri contains high-quality implementations of prominent unsupervised CBIR algorithms. As for the supported functionality, we have adopted an object-oriented approach, implementing each algorithm as a class template, while also providing free functions for simpler operations.  A list of supported methods is given as follows.

\subsection{Pre-processing Methods}\label{sec:pre-process}

\begin{itemize}[itemsep=0em, leftmargin=2em]
\item DirectResize (DR): Scaling the height and width of the image to the target size directly.
\item PadResize (PR): Scaling the longer side of the image to the target size and filling the remaining pixels with the mean values of ImageNet.
\item ShorterResize (SR): Scaling the shorter side of the image to the target size.
\item TwoFlip (TF): Returning the original image and the corresponding horizontally flipped image.
\item CenterCrop (CC): Cropping the image from its center region according to the given size.
\item TenCrop (TC): Cropping the original image and the flipping image from up down left right and center, respectively.
\end{itemize}

\vspace{-1em}
\begin{table*}[] 
\footnotesize
\newcommand{\tabincell}[2]{\begin{tabular}{@{}#1@{}}#2\end{tabular}}  
  \caption{Top-3 retrieval accuracy w.r.t. the corresponding searched configurations and the baseline of each dataset. Compared with the baseline consisting of default configurations, proper retrieval configurations can outperform it by a large margin.}
\vspace{-1em}
  \label{tab:object}
  \begin{tabular}{|c||c|c|c|c|c|c|c|c|c||c|}
    \hline
    \tabincell{c}{Dataset} & \tabincell{c}{Config.}  & \tabincell{c}{Pre-trained\\data}  & \tabincell{c}{Data\\aug} & \tabincell{c}{Backbone} & \tabincell{c}{Output\\layer} & \tabincell{c}{Aggregation} & \tabincell{c}{Dimension\\process} & \tabincell{c}{Feature\\enhance} & \tabincell{c}{Re-rank} & \tabincell{c}{mAP}\\
\hline
\hline
    \multirow{4}{*}{\textit{Oxford5k}}  & \multirow{3}{*}{\tabincell{c}{Search\\Configs.}} & ImageNet+Places & SR+CC & VGG16 & pool5 & GAP & $\ell_{2}$+$SVD_{w}$+$\ell_{2}$ & -- & $k$-reciprocal & \textbf{72.9 (+26.6)}\\
\cline{3-11} 
    						& & ImageNet+Places & DR & Res50 & pool5 & SPoC & $\ell_{2}$+$PCA_{w}$+$\ell_{2}$ & -- & $k$-reciprocal & 72.4\\ 
\cline{3-11} 
  				       	    & & ImageNet+Places & SR+CC & VGG16 & pool5 & GAP & $\ell_{2}$+$PCA_{w}$+$\ell_{2}$ & -- & $k$-reciprocal & 72.1\\   
\cline{2-11}     
				       & Baseline & ImageNet & SR+CC & Res50 & pool5 & GAP & -- & -- & -- & 46.3\\ \hline\hline
    \multirow{4}{*}{\textit{CUB-200}} & \multirow{3}{*}{\tabincell{c}{Search\\Configs.}} & ImageNet & SR+CC & Res50 & pool5 & SCDA & $\ell_{2}$+$PCA$+$\ell_{2}$ & - & $k$-reciprocal& \textbf{38.9 (+21.0)}\\
\cline{3-11} 
    						   & & ImageNet & SR+CC & Res50 & pool5 & GeM & $\ell_{2}$+$PCA$+$\ell_{2}$ & -- & $k$-reciprocal& 37.3\\
\cline{3-11} 
						   & & ImageNet & SR+CC & Res50 & pool5 & GMP & $\ell_{2}$+$PCA$+$\ell_{2}$ & -- & $k$-reciprocal& 37.2\\ 
						\cline{2-11} 
						   & Baseline & ImageNet & SR+CC & Res50 & pool5 & GAP & -- & -- & -- & 17.9\\ \hline\hline
    \multirow{4}{*}{\textit{Indoor}} & \multirow{3}{*}{\tabincell{c}{Search\\Configs.}} & Places & DR & Res50 & pool5 & CroW & $\ell_{2}$+$PCA$+$\ell_{2}$ & DBA & QE & \textbf{63.7 (+39.8)}\\
\cline{3-11}     				       
& & Places & DR & Res50 & pool5 & GAP & $\ell_{2}$+$PCA$+$\ell_{2}$ & DBA & QE & 63.5\\
\cline{3-11} 
				       & & Places & DR & Res50 & pool5 & GeM & $\ell_{2}$+$PCA$+$\ell_{2}$ & DBA & QE & 63.2\\ 
				\cline{2-11} 
				       & Baseline& ImageNet & SR+CC & Res50 & pool5 & GAP & -- & -- & -- & 23.9\\ \hline\hline
     \multirow{4}{*}{\textit{Caltech101}} & \multirow{3}{*}{\tabincell{c}{Search\\Configs.}} & ImageNet & PR & Res50 & pool5 & GMP & $\ell_{2}$+$PCA$+$\ell_{2}$ & DBA & QE+$k$-reciprocal& \textbf{86.4 (+19.3)}\\
\cline{3-11} 
					       & &ImageNet & PR & Res50 & pool5 & GeM & $\ell_{2}$+$PCA$+$\ell_{2}$ & DBA & QE+$k$-reciprocal& 86.1\\
					\cline{3-11} 
					       & & ImageNet & PR & Res50 & pool5 & SCDA & $\ell_{2}$+$PCA$+$\ell_{2}$ & DBA & QE+$k$-reciprocal& 86.1\\
					\cline{2-11} 
					       & Baseline & ImageNet & SR+CC & Res50 & pool5 & GAP & -- & -- & -- & 67.1\\ 
  \hline
\end{tabular}
\end{table*}

\subsection{Feature Representation Methods}
\begin{itemize}[itemsep=0em, leftmargin=2em]
\item GAP: Global average pooling.
\item GMP: Global max pooling.
\item R-MAC~\cite{tolias2015particular}: Calculating feature vectors based on the regional maximum activation of convolutions.
\item SPoC~\cite{babenko2015aggregating}: Assigning larger weights to the central descriptors during aggregation.
\item CroW~\cite{kalantidis2016cross}: A weighted pooling method for both spatial- and channel-wise.  
\item SCDA~\cite{wei2017selective}: Keeping useful deep descriptors based on the summation of feature map activations.
\item GeM~\cite{radenovic2018fine}: Exploiting the generalized mean to reserve the information of each channel.
\item PWA~\cite{xu2018unsupervised}: Aggregating the regional representations weighted by the selected part detectors' output.
\item PCB~\cite{sun2018beyond}: Outputting a convolutional descriptor consisting of several part-level features.
\end{itemize}


\subsection{Post-precessing Methods}

\begin{itemize}[itemsep=0em, leftmargin=2em]
\item SVD~\cite{golub1971singular}: Reducing feature dimension through singular value decomposition of matrix.
\item PCA~\cite{wold1987principal}: Projecting high-dimensional features into fewer informative dimensions.
\item DBA~\cite{arandjelovic2012three}: Every feature in the database is replaced with a weighted sum of the point's own value and those of its top $k$ nearest neighbors ($k$-NN).
\item QE~\cite{chum2007total}: Combining the retrieved top-$k$ nearest neighbors with the original query and doing another retrieval.
\item $k$-reciprocal~\cite{zhong2017re}: Encoding $k$-reciprocal nearest neighbors to enhance the accuracy of retrieval.
\end{itemize}

\begin{table}
\footnotesize
\setlength{\tabcolsep}{2.8pt}
\newcommand{\tabincell}[2]{\begin{tabular}{@{}#1@{}}#2\end{tabular}}  
  \caption{Inference speed comparisons of each retrieval stage.}
\vspace{-1em}
  \label{tab:speed}
  \begin{tabular}{|c||c|c|c||c|}
    \hline
    Stage & $\sharp$ images & Resolution & Backbone & Avg. time (ms/img)\\
    \hline
    \hline
    \multirow{2}{*}{Extraction} & \multirow{2}{*}{100} & \multirow{2}{*}{$224 \times 224$} & VGG16 & 7.16 \\ 
\cline{4-5} 
    					                                &  &  & Res50 & 7.72\\ \hline\hline
    \multirow{2}{*}{Indexing} & Query set: 100 & \multirow{2}{*}{$224 \times 224$} & VGG16 & 0.02\\ 
\cline{4-5} 
    					  & Gallery set: 100  & & Res50 & 0.03\\ 
  \hline
\end{tabular}
\end{table}

\section{Configuration Search Tool}

Since different algorithms used in each stage might have a significant impact on retrieval accuracy, we present the configuration search tool to help users to find the optimal retrieval configuration with various hyper-parameters.

As the same coding style as PyRetri, our configuration search tool is easy to read and deploy. The tool consists two components: the search space and search script. The search space is defined by the users through adding methods with hyper-parameters to a specified dict. Then, the search script completes the retrieval process based on all the configurations within the search space in an exhaustive way, saving the results automatically. 

Moreover, in order to help users analyze retrieval results easily, we also provide scripts to convert the results file into the \texttt{csv} format or filter the retrieval results according to the given key words.

\section{Applications and Evaluations}

Recently, deep learning based approaches are widely explored for unsupervised image retrieval, which have achieved satisfactory results and have been successfully applied to diverse multimedia and computer vision tasks like content-based image retrieval and person re-identification, etc.  We validate the effectiveness of PyRetri on the two tasks respectively through experiments on several benchmark datasets.

\subsection{Benchmark Datasets}
\begin{itemize}[itemsep=0em, leftmargin=2em]
\item Oxford5k~\cite{philbin2007object} collects crawling images from \texttt{Flickr} using the names of 11 different landmarks in Oxford, which is a representative landmark retrieval task.
\item CUB-200-2011~\cite{WahCUB_200_2011} contains photos of 200 bird species, which represents fine-grained image retrieval.
\item Indoor~\cite{quattoni2009recognizing} contains indoor scene images with 67 categories, representing for the scene retrieval/recognition task.
\item Caltech101~\cite{fei2004learning} consists pictures of objects belonging to 101 categories, standing for the generic image retrieval task.
\item Market-1501~\cite{zheng2015scalable} contains images taken on the Tsinghua campus under six camera viewpoints, which is the benchmark dataset for person re-identification.
\item DukeMTMC-reID~\cite{ristani2016performance} contains images captured by eight cameras, which is a more challenging person Re-ID dataset.
\end{itemize}

\subsection{Content-Based Image Retrieval}


For the configurations of content-based image retrieval, we pick up the following search factors for evaluation: data augmentation, backbone, aggregation methods and dimension process/reduction. The data augmentation operations include SR+CC, PR and DR (cf. Sec.~\ref{sec:pre-process}), aiming at finding the relationship between image integrity and the retrieval accuracy. All the aggregation methods are added to the search space in order to get comprehensive analyses. In addition, popularly used dimension process/reduction approaches are adopted for searching, such as $\ell_{2}$-normalization ($\ell_{2}$), PCA with whitening ($PCA_{w}$) or PCA without whitening ($PCA$) and SVD with whitening ($SVD_{w}$) or without whitening ($SVD$). After getting the top-3 best retrieval accuracy w.r.t. the searched configurations on each dataset, we further search for the optimal feature enhance and rerank operations. 

As reported in Table~\ref{tab:object}, proper retrieval configurations bring a significant improvement on the retrieval accuracy on these CBIR benchmark datasets. Also, the inference speed of each retrieval stage is shown in Table~\ref{tab:speed}. The averaged inference time per image by our PyRetri is less than 8ms.


\subsection{Person Re-Identification}
\label{sec:reid}
In addition to the general image retrieval task, we also apply PyRetri to person re-identification, which owns the retrieval workflow. 

Since the model for Re-ID tasks has been trained on the target dataset, we do not search for feature extraction operations and just evaluate the accuracy of PyRetri based on the open source pre-trained models. Since the developer does not give the model trained on DukeMTMC-reID, we train a model on it by ourselves by employing the widely used code~\cite{reid_baseline}, which is utilized as the baseline of the Re-ID experiments. 

As shown in Table~\ref{tab:reid}, our re-ID results are able to be the same as the results reported in the original implementation, which proves that our PyRetri is reliable. More importantly, by applying the retrieval configurations provided by PyRetri, our results outperform the baseline by a large margin, justifying the practical effectiveness of the library.

\begin{table}
\footnotesize
  \caption{Re-ID accuracy of benchmark datasets. Our implementation achieves accuracy on par with the reported results. With the optimal configuration provided by PyRetri, the results significantly boost.}
\vspace{-1em}
  \label{tab:reid}
  \begin{tabular}{|c||c||c|c|}
    \hline
    Dataset & Implementations & mAP & Recall@1\\
    \hline
\hline
    \multirow{3}{*}{\textit{Market-1501}} & Referenced impl.~\cite{reid_baseline} & 71.6 & 88.8\\
\cline{2-4} 
                                & Ours & 71.6 & 88.8\\
\cline{2-4} 
                                & Ours w. optimal config. & \textbf{84.8} & \textbf{90.4}\\ \hline\hline
     \multirow{3}{*}{\textit{DukeMTMC-reID}} & Referenced impl.~\cite{reid_baseline} & -- & --\\
\cline{2-4} 
                                & Ours & 62.5 & 80.4\\
\cline{2-4} 
                                & Ours w. optimal config. & \textbf{78.3} & \textbf{84.2}\\
  \hline
\end{tabular}
\end{table}

\section{Avaliability}
PyRetri is released under the license of Apache 2.0 and its source code is openly available at: \textbf{https://github.com/PyRetri/PyRetri}. We also provide extensive documentations and sampled projects for it. Contributions from the open-source community are welcome, via the GitHub issues/pull request mechanisms.

\section{Conclusions}

In this project, we have developed PyRetri, a software library that focuses on deep learning based unsupervised image retrieval. Our library unifies the pipeline of CBIR and provides convenient interfaces for each retrieval stage, which can be easily adopted for various multimedia application scenarios. With modular designs and object-oriented implementations, PyRetri is easy-to-use and easy-to-extend, which is suitable to be a codebase for other researchers. We hope that PyRetri can provide unique convenience to the deep learning and CBIR research community.
\bibliographystyle{ACM-Reference-Format}
\bibliography{sample-base}


\begin{thebibliography}{22}


\ifx \showCODEN    \undefined \def \showCODEN     #1{\unskip}     \fi
\ifx \showDOI      \undefined \def \showDOI       #1{#1}\fi
\ifx \showISBNx    \undefined \def \showISBNx     #1{\unskip}     \fi
\ifx \showISBNxiii \undefined \def \showISBNxiii  #1{\unskip}     \fi
\ifx \showISSN     \undefined \def \showISSN      #1{\unskip}     \fi
\ifx \showLCCN     \undefined \def \showLCCN      #1{\unskip}     \fi
\ifx \shownote     \undefined \def \shownote      #1{#1}          \fi
\ifx \showarticletitle \undefined \def \showarticletitle #1{#1}   \fi
\ifx \showURL      \undefined \def \showURL       {\relax}        \fi
\providecommand\bibfield[2]{#2}
\providecommand\bibinfo[2]{#2}
\providecommand\natexlab[1]{#1}
\providecommand\showeprint[2][]{arXiv:#2}

\bibitem[\protect\citeauthoryear{Arandjelovi{\'c} and
  Zisserman}{Arandjelovi{\'c} and Zisserman}{2012}]%
        {arandjelovic2012three}
\bibfield{author}{\bibinfo{person}{R. Arandjelovi{\'c}} {and}
  \bibinfo{person}{A. Zisserman}.} \bibinfo{year}{2012}\natexlab{}.
\newblock \showarticletitle{Three things everyone should know to improve object
  retrieval}. In \bibinfo{booktitle}{\emph{CVPR}}. \bibinfo{pages}{2911--2918}.
\newblock


\bibitem[\protect\citeauthoryear{Babenko and Lempitsky}{Babenko and
  Lempitsky}{2015}]%
        {babenko2015aggregating}
\bibfield{author}{\bibinfo{person}{A. Babenko} {and} \bibinfo{person}{V.
  Lempitsky}.} \bibinfo{year}{2015}\natexlab{}.
\newblock \showarticletitle{Aggregating deep convolutional features for image
  retrieval}. In \bibinfo{booktitle}{\emph{ICCV}}. \bibinfo{pages}{1269--1277}.
\newblock


\bibitem[\protect\citeauthoryear{Chum, Philbin, Sivic, Isard, and
  Zisserman}{Chum et~al\mbox{.}}{2007}]%
        {chum2007total}
\bibfield{author}{\bibinfo{person}{O. Chum}, \bibinfo{person}{J. Philbin},
  \bibinfo{person}{J. Sivic}, \bibinfo{person}{M. Isard}, {and}
  \bibinfo{person}{A. Zisserman}.} \bibinfo{year}{2007}\natexlab{}.
\newblock \showarticletitle{Total recall: Automatic query expansion with a
  generative feature model for object retrieval}. In
  \bibinfo{booktitle}{\emph{ICCV}}. \bibinfo{pages}{1--8}.
\newblock


\bibitem[\protect\citeauthoryear{Fei-Fei, Fergus, and Perona}{Fei-Fei
  et~al\mbox{.}}{2004}]%
        {fei2004learning}
\bibfield{author}{\bibinfo{person}{L. Fei-Fei}, \bibinfo{person}{R. Fergus},
  {and} \bibinfo{person}{P. Perona}.} \bibinfo{year}{2004}\natexlab{}.
\newblock \showarticletitle{Learning generative visual models from few training
  examples: An incremental bayesian approach tested on 101 object categories}.
  In \bibinfo{booktitle}{\emph{CVPR workshop}}. \bibinfo{pages}{178--186}.
\newblock


\bibitem[\protect\citeauthoryear{Girshick}{Girshick}{2018}]%
        {yacs}
\bibfield{author}{\bibinfo{person}{R. Girshick}.}
  \bibinfo{year}{2018}\natexlab{}.
\newblock \bibinfo{title}{YACS}.
\newblock \bibinfo{howpublished}{Website}.
\newblock
\newblock
\shownote{\url{https://github.com/rbgirshick/yacs}.}


\bibitem[\protect\citeauthoryear{Golub and Reinsch}{Golub and Reinsch}{1970}]%
        {golub1971singular}
\bibfield{author}{\bibinfo{person}{G.H. Golub} {and} \bibinfo{person}{C.
  Reinsch}.} \bibinfo{year}{1970}\natexlab{}.
\newblock \showarticletitle{Singular value decomposition and least squares
  solutions}.
\newblock \bibinfo{journal}{\emph{Numer. Math.}} \bibinfo{volume}{14},
  \bibinfo{number}{5} (\bibinfo{year}{1970}), \bibinfo{pages}{403--420}.
\newblock


\bibitem[\protect\citeauthoryear{Kalantidis, Mellina, and Osindero}{Kalantidis
  et~al\mbox{.}}{2016}]%
        {kalantidis2016cross}
\bibfield{author}{\bibinfo{person}{Y. Kalantidis}, \bibinfo{person}{C.
  Mellina}, {and} \bibinfo{person}{S. Osindero}.}
  \bibinfo{year}{2016}\natexlab{}.
\newblock \showarticletitle{Cross-dimensional weighting for aggregated deep
  convolutional features}. In \bibinfo{booktitle}{\emph{ECCV workshop}}.
  \bibinfo{pages}{685--701}.
\newblock


\bibitem[\protect\citeauthoryear{Paszke, Gross, Massa, Lerer, Bradbury, Chanan,
  Killeen, Lin, Gimelshein, Antiga, Desmaison, Köpf, Yang, DeVito, Raison,
  Tejani, Chilamkurthy, Steiner, Fang, Bai, and Chintala}{Paszke
  et~al\mbox{.}}{2019}]%
        {paszke2019pytorch}
\bibfield{author}{\bibinfo{person}{A. Paszke}, \bibinfo{person}{S. Gross},
  \bibinfo{person}{F. Massa}, \bibinfo{person}{A. Lerer}, \bibinfo{person}{J.
  Bradbury}, \bibinfo{person}{G. Chanan}, \bibinfo{person}{T. Killeen},
  \bibinfo{person}{Z. Lin}, \bibinfo{person}{N. Gimelshein},
  \bibinfo{person}{L. Antiga}, \bibinfo{person}{A. Desmaison},
  \bibinfo{person}{A. Köpf}, \bibinfo{person}{E. Yang}, \bibinfo{person}{Z.
  DeVito}, \bibinfo{person}{M. Raison}, \bibinfo{person}{A. Tejani},
  \bibinfo{person}{S. Chilamkurthy}, \bibinfo{person}{B. Steiner},
  \bibinfo{person}{L. Fang}, \bibinfo{person}{J. Bai}, {and}
  \bibinfo{person}{S. Chintala}.} \bibinfo{year}{2019}\natexlab{}.
\newblock \showarticletitle{PyTorch: An imperative style, high-performance deep
  learning library}. In \bibinfo{booktitle}{\emph{NeurIPS}}.
  \bibinfo{pages}{8024--8035}.
\newblock


\bibitem[\protect\citeauthoryear{Philbin, Chum, Isard, Sivic, and
  Zisserman}{Philbin et~al\mbox{.}}{2007}]%
        {philbin2007object}
\bibfield{author}{\bibinfo{person}{J. Philbin}, \bibinfo{person}{O. Chum},
  \bibinfo{person}{M. Isard}, \bibinfo{person}{J. Sivic}, {and}
  \bibinfo{person}{A. Zisserman}.} \bibinfo{year}{2007}\natexlab{}.
\newblock \showarticletitle{Object retrieval with large vocabularies and fast
  spatial matching}. In \bibinfo{booktitle}{\emph{CVPR}}.
  \bibinfo{pages}{1--8}.
\newblock


\bibitem[\protect\citeauthoryear{Quattoni and Torralba}{Quattoni and
  Torralba}{2009}]%
        {quattoni2009recognizing}
\bibfield{author}{\bibinfo{person}{A. Quattoni} {and} \bibinfo{person}{A.
  Torralba}.} \bibinfo{year}{2009}\natexlab{}.
\newblock \showarticletitle{Recognizing indoor scenes}. In
  \bibinfo{booktitle}{\emph{CVPR}}. \bibinfo{pages}{413--420}.
\newblock


\bibitem[\protect\citeauthoryear{Radenovi{\'c}, Tolias, and Chum}{Radenovi{\'c}
  et~al\mbox{.}}{2019}]%
        {radenovic2018fine}
\bibfield{author}{\bibinfo{person}{F. Radenovi{\'c}}, \bibinfo{person}{G.
  Tolias}, {and} \bibinfo{person}{O. Chum}.} \bibinfo{year}{2019}\natexlab{}.
\newblock \showarticletitle{Fine-tuning CNN image retrieval with no human
  annotation}.
\newblock \bibinfo{journal}{\emph{IEEE TPAMI}} \bibinfo{volume}{41},
  \bibinfo{number}{7} (\bibinfo{year}{2019}), \bibinfo{pages}{1655--1668}.
\newblock


\bibitem[\protect\citeauthoryear{Ristani, Solera, Zou, Cucchiara, and
  Tomasi}{Ristani et~al\mbox{.}}{2016}]%
        {ristani2016performance}
\bibfield{author}{\bibinfo{person}{E. Ristani}, \bibinfo{person}{F. Solera},
  \bibinfo{person}{R. Zou}, \bibinfo{person}{R. Cucchiara}, {and}
  \bibinfo{person}{C. Tomasi}.} \bibinfo{year}{2016}\natexlab{}.
\newblock \showarticletitle{Performance measures and a data set for
  multi-target, multi-camera tracking}. In \bibinfo{booktitle}{\emph{ECCV
  workshop}}. \bibinfo{pages}{17--35}.
\newblock


\bibitem[\protect\citeauthoryear{Sun, Zheng, Yang, Tian, and Wang}{Sun
  et~al\mbox{.}}{2018}]%
        {sun2018beyond}
\bibfield{author}{\bibinfo{person}{Y. Sun}, \bibinfo{person}{L. Zheng},
  \bibinfo{person}{Y. Yang}, \bibinfo{person}{Q. Tian}, {and}
  \bibinfo{person}{S. Wang}.} \bibinfo{year}{2018}\natexlab{}.
\newblock \showarticletitle{Beyond part models: Person retrieval with refined
  part pooling}. In \bibinfo{booktitle}{\emph{ECCV}}.
  \bibinfo{pages}{501--518}.
\newblock


\bibitem[\protect\citeauthoryear{Tolias, Sicre, and J{\'e}gou}{Tolias
  et~al\mbox{.}}{2016}]%
        {tolias2015particular}
\bibfield{author}{\bibinfo{person}{G. Tolias}, \bibinfo{person}{R. Sicre},
  {and} \bibinfo{person}{H. J{\'e}gou}.} \bibinfo{year}{2016}\natexlab{}.
\newblock \showarticletitle{Particular object retrieval with integral
  max-pooling of CNN activations}. In \bibinfo{booktitle}{\emph{ICLR}}.
  \bibinfo{pages}{1--12}.
\newblock


\bibitem[\protect\citeauthoryear{Wah, Branson, Welinder, Perona, and
  Belongie}{Wah et~al\mbox{.}}{2011}]%
        {WahCUB_200_2011}
\bibfield{author}{\bibinfo{person}{C. Wah}, \bibinfo{person}{S. Branson},
  \bibinfo{person}{P. Welinder}, \bibinfo{person}{P. Perona}, {and}
  \bibinfo{person}{S. Belongie}.} \bibinfo{year}{2011}\natexlab{}.
\newblock \bibinfo{booktitle}{\emph{{The Caltech-UCSD Birds-200-2011
  Dataset}}}.
\newblock \bibinfo{type}{{T}echnical {R}eport} CNS-TR-2011-001.
  \bibinfo{institution}{California Institute of Technology}.
\newblock


\bibitem[\protect\citeauthoryear{Wan, Wang, Hoi, Wu, Zhu, Zhang, and Li}{Wan
  et~al\mbox{.}}{2014}]%
        {cbirsurvey}
\bibfield{author}{\bibinfo{person}{J. Wan}, \bibinfo{person}{D. Wang},
  \bibinfo{person}{S.~C.{-}H. Hoi}, \bibinfo{person}{P. Wu},
  \bibinfo{person}{J. Zhu}, \bibinfo{person}{Y. Zhang}, {and}
  \bibinfo{person}{J. Li}.} \bibinfo{year}{2014}\natexlab{}.
\newblock \showarticletitle{Deep Learning for Content-Based Image Retrieval:
  {A} Comprehensive Study}. In \bibinfo{booktitle}{\emph{{ACM}, {MM}}}.
  \bibinfo{pages}{157--166}.
\newblock


\bibitem[\protect\citeauthoryear{Wei, Luo, Wu, and Zhou}{Wei
  et~al\mbox{.}}{2017}]%
        {wei2017selective}
\bibfield{author}{\bibinfo{person}{X.-S. Wei}, \bibinfo{person}{J.-H. Luo},
  \bibinfo{person}{J. Wu}, {and} \bibinfo{person}{Z.-H. Zhou}.}
  \bibinfo{year}{2017}\natexlab{}.
\newblock \showarticletitle{Selective convolutional descriptor aggregation for
  fine-grained image retrieval}.
\newblock \bibinfo{journal}{\emph{IEEE TIP}} \bibinfo{volume}{26},
  \bibinfo{number}{6} (\bibinfo{year}{2017}), \bibinfo{pages}{2868--2881}.
\newblock


\bibitem[\protect\citeauthoryear{Wold, Esbensen, and Geladi}{Wold
  et~al\mbox{.}}{1987}]%
        {wold1987principal}
\bibfield{author}{\bibinfo{person}{S. Wold}, \bibinfo{person}{K. Esbensen},
  {and} \bibinfo{person}{P. Geladi}.} \bibinfo{year}{1987}\natexlab{}.
\newblock \showarticletitle{Principal component analysis}.
\newblock \bibinfo{journal}{\emph{Chemometrics and intelligent laboratory
  systems}} \bibinfo{volume}{2}, \bibinfo{number}{1-3} (\bibinfo{year}{1987}),
  \bibinfo{pages}{37--52}.
\newblock


\bibitem[\protect\citeauthoryear{Xu, Shi, Qi, Wang, and Xiao}{Xu
  et~al\mbox{.}}{2018}]%
        {xu2018unsupervised}
\bibfield{author}{\bibinfo{person}{J. Xu}, \bibinfo{person}{C. Shi},
  \bibinfo{person}{C. Qi}, \bibinfo{person}{C. Wang}, {and} \bibinfo{person}{B.
  Xiao}.} \bibinfo{year}{2018}\natexlab{}.
\newblock \showarticletitle{Unsupervised part-based weighting aggregation of
  deep convolutional features for image retrieval}. In
  \bibinfo{booktitle}{\emph{AAAI}}. \bibinfo{pages}{7436--7443}.
\newblock


\bibitem[\protect\citeauthoryear{Zheng, Shen, Tian, Wang, Wang, and Tian}{Zheng
  et~al\mbox{.}}{2015}]%
        {zheng2015scalable}
\bibfield{author}{\bibinfo{person}{L. Zheng}, \bibinfo{person}{L. Shen},
  \bibinfo{person}{L. Tian}, \bibinfo{person}{S. Wang}, \bibinfo{person}{J.
  Wang}, {and} \bibinfo{person}{Q. Tian}.} \bibinfo{year}{2015}\natexlab{}.
\newblock \showarticletitle{Scalable person re-identification: A benchmark}. In
  \bibinfo{booktitle}{\emph{ICCV}}. \bibinfo{pages}{1116--1124}.
\newblock


\bibitem[\protect\citeauthoryear{Zheng}{Zheng}{2018}]%
        {reid_baseline}
\bibfield{author}{\bibinfo{person}{Z. Zheng}.} \bibinfo{year}{2018}\natexlab{}.
\newblock
\newblock
\newblock
\shownote{\url{https://github.com/layumi/Person_reID_baseline_pytorch}.}


\bibitem[\protect\citeauthoryear{Zhong, Zheng, Cao, and Li}{Zhong
  et~al\mbox{.}}{2017}]%
        {zhong2017re}
\bibfield{author}{\bibinfo{person}{Z. Zhong}, \bibinfo{person}{L. Zheng},
  \bibinfo{person}{D. Cao}, {and} \bibinfo{person}{S. Li}.}
  \bibinfo{year}{2017}\natexlab{}.
\newblock \showarticletitle{Re-ranking person re-identification with
  k-reciprocal encoding}. In \bibinfo{booktitle}{\emph{CVPR}}.
  \bibinfo{pages}{1318--1327}.
\newblock


\end{thebibliography}

\appendix

\end{document}